\documentclass[aps,pra,showpacs,amssymb,superscriptaddress,footinbib,twocolumn]{revtex4-1}
\usepackage[ansinew]{inputenc}
\usepackage{bbm}
\usepackage{bm}
\usepackage{amsbsy}
\usepackage{amsthm}
\usepackage{amssymb}
\usepackage{amsfonts}
\usepackage{amsmath}
\usepackage{dsfont} 
\usepackage{graphicx} 
\usepackage{epsfig}
\usepackage{epstopdf}
\usepackage{dsfont}
\usepackage{color}
\usepackage[colorlinks]{hyperref}
\usepackage[figure,table]{hypcap}
\usepackage{enumerate}
\hypersetup{
	bookmarksnumbered,
	pdfstartview={FitH},
	citecolor={darkgreen},
	linkcolor={darkred},
	urlcolor={darkblue},
	pdfpagemode={UseOutlines}}
\definecolor{darkgreen}{RGB}{50,190,50}
\definecolor{darkblue}{RGB}{0,0,190}
\definecolor{darkred}{RGB}{238,0,0}
\usepackage{soul}
\newcommand{\pr}{^{\prime}}
\newcommand{\ket}[1]{\ensuremath{\left|{#1}\right\rangle}}
\newcommand{\ketsub}[2]{\ensuremath{\left|{#1}\hspace*{1pt}\right\rangle_{\hspace*{-1pt}#2}}}
\newcommand{\bra}[1]{\ensuremath{\left\langle{#1}\right|}}

\newcommand{\djj}{d\kern-0.4em\char"16\kern-0.1em}

\begin{document}

\title{
Implementing quantum control for unknown subroutines}
\author{Nicolai Friis}
\email{nicolai.friis@uibk.ac.at}
\affiliation{
Institute for Quantum Optics and Quantum Information,
Austrian Academy of Sciences,
Technikerstra{\ss}e 21a,
A-6020 Innsbruck,
Austria}

\author{Vedran Dunjko}
\email{vedran.dunjko@uibk.ac.at}
\affiliation{
Institute for Quantum Optics and Quantum Information,
Austrian Academy of Sciences,
Technikerstra{\ss}e 21a,
A-6020 Innsbruck,
Austria}
\affiliation{
Institute for Theoretical Physics, University of Innsbruck,
Technikerstra{\ss}e 25,
A-6020 Innsbruck,
Austria}
\affiliation{Laboratory of Evolutionary Genetics, Division of Molecular Biology,
Ru{\djj}er Bo\v{s}kovi{\'c} Institute, Bijeni\v{c}ka cesta 54, HR-10000 Zagreb, Croatia}

\author{Wolfgang D{\"u}r}
\email{wolfgang.duer@uibk.ac.at}
\affiliation{
Institute for Theoretical Physics, University of Innsbruck,
Technikerstra{\ss}e 25,
A-6020 Innsbruck,
Austria}

\author{Hans J.~Briegel}
\email{hans.briegel@uibk.ac.at}
\affiliation{
Institute for Quantum Optics and Quantum Information,
Austrian Academy of Sciences,
Technikerstra{\ss}e 21a,
A-6020 Innsbruck,
Austria}
\affiliation{
Institute for Theoretical Physics, University of Innsbruck,
Technikerstra{\ss}e 25,
A-6020 Innsbruck,
Austria}
\date{\today}
\begin{abstract}
We present setups for the practical realization of adding control to unknown subroutines, supplementing the existing quantum optical scheme for black-box control with a counterpart for the quantum control of the ordering of sequences of operations. We also provide schemes to realize either task using trapped ions. These practical circumventions of recent no-go theorems are based on existing technologies. We argue that the possibility to add control to unknown operations in practice is a common feature of many physical systems. Based on the proposed implementations we discuss the apparent contradictions between theory and practice. 
\end{abstract}

\maketitle
\section{Introduction}
\begin{figure}[ht!]
\includegraphics[width=0.42\textwidth]{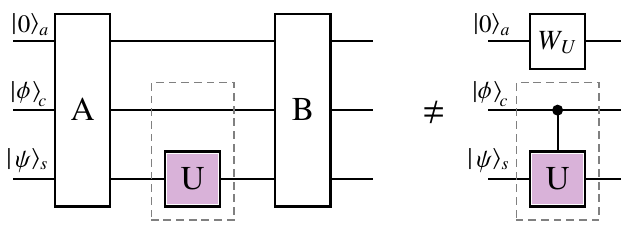}
\caption{\label{fig:controlled U no go}
\textbf{Circuit representation of no-go theorem}. As shown in Refs.~\cite{AraujoFeixCostaBrukner2013,ThompsonGuModiVedral2013}, it is not possible to construct a quantum circuit that deterministically realizes $\operatorname{ctrl}\textendash U$ for an unknown unitary~$U$, by using gates~$A$ and~$B$ that are independent of~$U$, even if the ancilla can undergo arbitrary transformations~$W_{U}$. From bottom to top the lines denote the system, the control qubit, and the ancilla. The mathematical meanings of the subcircuits in the dashed boxes, while
different on both sides, are nonetheless independent of the remaining circuits.\vspace*{-3mm}}
\end{figure}

\vspace*{-1mm}
Recently, there has been renewed interest in the problem of controlling unknown operations~\cite{AraujoFeixCostaBrukner2013,ThompsonGuModiVedral2013}, which had previously been addressed in an experiment~\cite{ZhouRalphKalasuwanZhangPeruzzoLanyonOBrien2011}, and the related problem of controlling their order~\cite{ChiribellaDArianoPerinottiValiron2013}, within the circuit model of quantum computation (see, e.g., Ref.~\cite{NielsenChuang2000} for an introduction). From a computational point of view, it is desirable to equip the quantum computer with a conditional control flow mechanism\textemdash a generic circuit realizing an ``\emph{if}" clause that takes as its input a number of unknown gates and implements these conditionally on the state of a control qubit\textemdash which is ubiquitous in essentially all classical programming languages. For instance, if an unknown unitary~$U$ is supplied, the naive expectation is that it can be inserted into a prefabricated circuit that is independent of~$U$, which performs the operation $\operatorname{ctrl}\textendash U$. However, various \emph{no-control theorems}~\cite{AraujoFeixCostaBrukner2013,ThompsonGuModiVedral2013} show that such a construction is not allowed by the mathematical structure of quantum mechanics (see Fig.~\ref{fig:controlled U no go}). Similarly, given single uses of two unknown operations, $U_{f}$ and $U_{g}$, it was shown in Ref.~\cite{ChiribellaDArianoPerinottiValiron2013} that no generic quantum circuit exists that allows for the deterministic control of the order, $U_{g}U_{f}$ or $U_{f}U_{g}$, of these operations. The former case, adding control to unknown operations, may be achieved in a restricted scenario if the control is classical~\cite{AraujoFeixCostaBrukner2013}. Interestingly, for sequence control even such a restriction does not resolve the problem~\cite{ChiribellaDArianoPerinottiValiron2013}.\\

\vspace*{-3.5mm}
A theorem cannot be broken. It thus seems that a fundamental impasse has been reached. In particular, one may take the view that these results limit the generally advocated superiority of quantum computers\\ 

\begin{figure}
\includegraphics[width=0.4\textwidth]{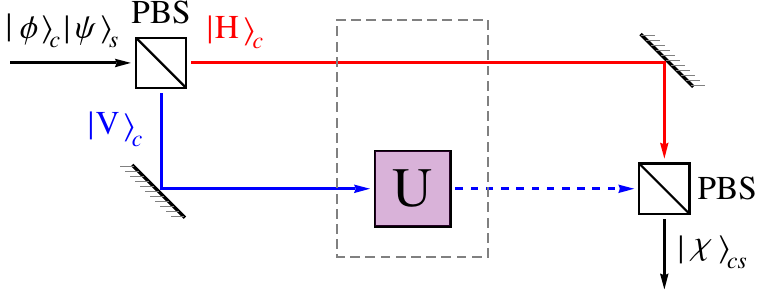}
\caption{\label{fig:cheating scheme control U}
\textbf{Quantum control of a single device}. In the scheme of Ref.~\cite{ZhouRalphKalasuwanZhangPeruzzoLanyonOBrien2011}, also discussed in Ref.~\cite{AraujoFeixCostaBrukner2013}, a single photon in the state $\ketsub{\phi}{c}\ketsub{\psi}{s}$, where the polarization state $\ketsub{\phi}{c}=\alpha\ketsub{\mathrm{H}}{c}+\beta\ketsub{\mathrm{V}}{c}$ serves as the control qubit and $\ketsub{\psi}{s}$ denotes the state of an additional degree of freedom, for instance, orbital angular momentum, is inserted into an interferometer. The polarizing beam splitters (PBSs) are configured to transmit and reflect horizontal and vertical polarization, respectively. The unknown unitary~$U$ is assumed to act only on the system $\ketsub{\psi}{s}$. After recombination at the second PBS the initial state has been transformed to $\ketsub{\protect\raisebox{1pt}{\textit{$\chi$}}}{cs}=\alpha\ketsub{\mathrm{H}}{c}\ketsub{\psi}{s}+\beta\ketsub{\mathrm{V}}{c}U\ketsub{\psi}{s}$.\vspace*{-3mm}}
\end{figure}

\noindent
\textemdash for instance, in their flexibility to handle black-box algorithms~\cite{ThompsonGuModiVedral2013}. However, if the underlying assumptions are changed, this impasse can be circumvented. In spite of the no-go theorems, it has been demonstrated~\cite{ZhouRalphKalasuwanZhangPeruzzoLanyonOBrien2011} that it is indeed possible to add control to a physical mechanism realizing an unknown unitary. More specifically, given a single physical device that performs an unknown unitary on an individual qubit, this device can be inserted into a setup that performs this operation conditionally on the state of another qubit. This is achieved by exploiting additional dimensions or degrees of freedom of the physical setup with respect to the mathematical framework assumed in the theorem. An 
example in a quantum optical setting, which has been discussed in Ref.~\cite{ZhouRalphKalasuwanZhangPeruzzoLanyonOBrien2011,AraujoFeixCostaBrukner2013}, is shown in Fig.~\ref{fig:cheating scheme control U}.\\

\vspace*{-2.5mm}
In this paper we introduce a novel scheme using trapped ions that allows one to add quantum control to unknown unitaries (see Fig.~\ref{fig:control with ions}). Having done this, we present two setups\textemdash with trapped ions and photons, respectively\textemdash that can realize the quantum-controlled switch of two operations for which we require only one ``use" of (``call" to) each unknown physical operation. We emphasize the single \emph{``use"} here, since its interpretation constitutes the pillar of the no-go result. As we shall discuss, practically adding control to unknown unitaries or the ordering of sequences thereof is not limited to these implementations and can also be achieved in other setups, e.g., in optical lattices~\cite{BriegelCalarcoJakschCiracZoller2000}, to name but one option. Indeed, whenever only a restricted part of the available Hilbert space is used in the physical realization of the qubits, options for adding control to unknown operations and their ordering may arise. In fact, the information about the subspace that the operation is \emph{not} acting on leaves the leeway for adding control\textemdash the operation is not entirely unknown.

The apparent ``loopholes" provided by practical settings circumventing the described no-go restrictions may seem to question the completeness of abstract quantum circuits. For instance, one may ask if quantum circuits are able to sufficiently represent the practical possibilities in the laboratory. Through an analysis of the structure of controlled processes and their representations we reach the conclusion that indeed every practical setup of quantum operations that can be realized in an experiment, e.g., on an optical table or with trapped atoms, can be represented as a quantum circuit. However, the schematic diagrams used for the former, albeit reminiscent of circuit diagrams, may not abide by the rules of the circuit model. Therefore, the correct circuit representation of a schematic may not resemble it visually. The practical implementations presented next will guide us to further clarify the relation between physical realizations, mathematical formulations, and graphical representations.\\
\vspace*{-9mm}

\section{Ion implementation}
Let us start by presenting a feasible scheme for black-box control using trapped ions. In the simplest\\

\vspace*{-3mm}
\begin{figure}[ht!]
\includegraphics[width=0.33\textwidth]{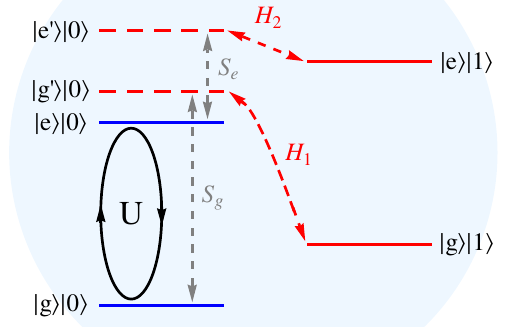}
\caption{\label{fig:control with ions}
\textbf{Controlling unknown unitaries with ions}. The collective vibrational mode of two ions is used to create two submanifolds of electronic energy levels within each ion. The excitation of the vibration introduces a fixed energy shift between the qubit levels $\ket{g}\hspace*{-1pt}\ket{1}$ and $\ket{e}\hspace*{-1pt}\ket{1}$ on the right, with respect to $\ket{g}\hspace*{-1pt}\ket{0}$ and $\ket{e}\hspace*{-1pt}\ket{0}$ on the left. The auxiliary levels $\ket{g\pr}$ and $\ket{e\pr}$ (horizontal dashed lines) are chosen such that the hiding pulses~$H_{1}$ and~$H_{2}$, and the $\sigma_{x}$ pulses~$S_{g}$ and~$S_{e}$ have different frequencies and are off resonance with the qubit transitions between $\ket{g}\hspace*{-1pt}\ket{n}$ and $\ket{e}\hspace*{-1pt}\ket{n}$ $(n=1,2,\ldots)$.\vspace*{-3mm}}
\end{figure}

\noindent
case two ions are confined in a linear trap in which they are laser cooled to the ground state~$\ket{0}$ of one of their common axial vibrational modes. Two electronic levels,~$\ket{g}_{i}$ and~$\ket{e}_{i}$, where $i=1,2$ labels the ions, are chosen to serve as the qubit levels (see Fig.~\ref{fig:control with ions}), such that the initial state is~$\bigl(\alpha\ket{g}_{1}+\beta\ket{e}_{1}\bigl)\ket{\psi}_{2}\hspace*{-1pt}\ket{0}$. The unknown unitary~$U$ is applied by a single laser pulse on ion~$2$ that resonantly drives the transition between~$\ket{g}_{2}$ and~$\ket{e}_{2}$. Conditioning~$U$ on the electronic state of the first ion can then be achieved with the following simple steps:
\begin{enumerate}[\hspace*{-2mm}(i)]
\item{\label{step i}Using the Cirac-Zoller method~\cite{CiracZoller1995}, an appropriately
blue-detuned laser is used to swap the qubit state from ion~$1$ to the common vibrational degree of freedom. The transition occurs exclusively from $\ket{g}_{1}\hspace*{-1pt}\ket{0}$ to $\ket{e}_{1}\hspace*{-1pt}\ket{1}$, such that the state is transformed from $\bigl(\alpha\ket{g}_{1}+\beta\ket{e}_{1}\bigl)\ket{\psi}_{2}\hspace*{-1pt}\ket{0}$ to $\ket{e}_{1}\ket{\psi}_{2}\hspace*{-1pt}\bigl(\alpha\ket{1}+\beta\ket{0}\hspace*{-1pt}\bigl)$.
\vspace*{-2.0mm}}
\item{\label{step ii}Two red-detuned hiding pulses~\cite{Barreiro-Blatt2011}, indicated by~$H_{1}$ and $H_{2}$ in Fig.~\ref{fig:control with ions}, on the second ion are used to transfer the populations of $\ket{g}_{2}\hspace*{-1pt}\ket{1}$ and $\ket{e}_{2}\hspace*{-1pt}\ket{1}$ to the auxiliary levels $\ket{g\pr}_{2}\hspace*{-1pt}\ket{0}$ and $\ket{e\pr}_{2}\hspace*{-1pt}\ket{0}$, respectively, which yields $\ket{e}_{1}\bigl(\alpha\ket{\psi\pr}_{2}\hspace*{-1pt}\ket{1}+\beta\ket{\psi}_{2}\hspace*{-1pt}\ket{0}\hspace*{-1pt}\bigl)\,$.
    \vspace*{-2.5mm}}
\item{\label{step iii}The pulse corresponding to the unknown unitary~$U$ is applied to ion~$2$, where only the transition between $\ket{g}_{2}\hspace*{-1pt}\ket{0}$ and $\ket{e}_{2}\hspace*{-1pt}\ket{0}$ is resonant, resulting in the state $\ket{e}_{1}\bigl(\alpha\ket{\psi\pr}_{2}\hspace*{-1pt}\ket{1}+\beta~U\ket{\psi}_{2}\hspace*{-1pt}\ket{0}\hspace*{-1pt}\bigl)\,$.
    \vspace*{-2.5mm}}
\item{\label{step iv}The hiding pulses~$H_{1}$ and $H_{2}$ are used to reverse the process of step~(\ref{step ii}), such that $\ket{\psi\pr}_{2}\rightarrow\ket{\psi}_{2}$\,.
    \vspace*{-2.5mm}}
\item{\label{step v}Another laser pulse on the first ion swaps the control back from the vibrational mode, $\ket{e}_{1}\hspace*{-1pt}\ket{1}\rightarrow\ket{g}_{1}\hspace*{-1pt}\ket{0}\,$, leaving the final state $\bigl(\alpha\ket{g}_{1}\ket{\psi}_{2}+\beta\ket{e}_{1}U\ket{\psi}_{2}\hspace*{-0.5pt}\bigr)\hspace*{-0.5pt}\ket{0}$, which completes the scheme.}
\end{enumerate}
All of the individual steps of our proposal can be considered to be part of the repertoire of established ion trap laboratories (see, e.g., Refs.~\cite{SchmidtKaler-Blatt2003,Barreiro-Blatt2011}). Hence, our setup is well within reach of state-of-the-art technology. We want to point out two main conceptual differences between the optical and ionic realizations. First, our scheme with trapped ions realizes the control qubit on a separate physical system (ion~$1$). This is advantageous in a scalable computational architecture and it can also in principle be used to straightforwardly control operations on an arbitrary number of ions via additional modes of their collective motion. Second, although the hiding pulses take on the role of the beam splitters in the optical setup, the additional degrees of freedom are not spatially delocalized as in a typical interferometer or in the ``movable wires," as suggested in Ref.~\cite{ChiribellaDArianoPerinottiValiron2013}.

\section{Black-box control}
With these considerations in mind, let us return to the seemingly paradox issue of explaining how control can be added to an unknown unitary in spite of the no-go results in Refs.~\cite{AraujoFeixCostaBrukner2013,ThompsonGuModiVedral2013}. As pointed out in Ref.~\cite{AraujoFeixCostaBrukner2013}, the operation~$U$ on the left-hand side of Fig.~\ref{fig:controlled U no go} acts on a subsystem~$\mathcal{H}_{s}$, while in Fig.~\ref{fig:cheating scheme control U} the device also labeled~$U$, in a slight abuse of notation, acts on a linear subspace of~$\mathcal{H}_{c}\otimes\mathcal{H}_{s}$, rather than a subsystem (particular degree of freedom). If we associate $\ketsub{\mathrm{H}}{c}$ and $\ketsub{\mathrm{V}}{c}$ with the eigenstates of~$\sigma_{z}$, we may write the controlled operation within the dashed box in Fig.~\ref{fig:cheating scheme control U} as $\mathds{1}_{d}\oplus U$, where~$\mathds{1}_{d}$ is the identity operator in $d=\dim(\mathcal{H}_{s})$ dimensions. In essence, this simply means that the setup in Fig.~\ref{fig:cheating scheme control U} indeed realizes the circuit shown on the right-hand side of Fig.~\ref{fig:controlled U no go}, despite the visual similarity with the left-hand side of Fig.~\ref{fig:controlled U no go}. The crucial observation here lies in the comparison of the dashed boxes in both figures. The symbols in the circuit diagram have a precise mathematical meaning\textemdash they are direct translations of mathematical expressions. In particular, the wires in the circuit model represent fixed subsystems. Conversely, the dashed box and the box labeled~$U$ in the schematic diagram of Fig.~\ref{fig:cheating scheme control U} have \emph{no} unambiguous 
meaning independently of the \emph{context}. The remainder of the schematic, along with information often not represented graphically, provide this context, which is needed to determine the mathematical representation of the device. In the case of Fig.~\ref{fig:cheating scheme control U} the information that one is dealing with an optical interferometer with a \emph{single} photon input is essential. The context thus consists of additional information about the relevant Fock space and how~$U$ acts on it. To illustrate this, consider the situation where all equipment outside the dashed box is removed, and one photon each is sent through the upper~(u) and lower~(l) path to generate $\mathds{1}_{\mathrm{u}}\otimes U_{\mathrm{l}}$. In other words, the no-go theorem does not apply, because the ``wires" in the schematic figure are not wires in the sense of the circuit diagram\textemdash they do not carry exactly one qubit each. For a circuit diagram symbol, the wires extending from it form an integral part of its meaning\textemdash they promise a particular input-output Hilbert space. For isolated symbols in schematic diagrams this is not the case.

Nonetheless, if the requirements of the circuit diagram are met, the no-go theorem applies. This would be the case, e.g., if the device performing the operation~$U$ in Fig.~\ref{fig:cheating scheme control U} contains an internal non-demolition photon number measurement. An interferometer setting modified in this way projects into either~$\ketsub{\mathrm{H}}{c}\ketsub{\psi}{s}$ or $\ketsub{\mathrm{V}}{c}U\ketsub{\psi}{s}$. In practice, such considerations will be of interest in adversarial scenarios related to quantum one-time programs~\cite{BroadbentGutoskiStebila2013}, that is, when one party grants limited access to the use of the $U$-device and wants to ensure that it is being employed according to the restrictions of the theorem. In this case, the circumvention of the no-go result becomes a competition between the user, exploiting additional degrees of freedom of the physical implementation, and the provider, who aims to countermand these attempts by more sophisticated devices\textemdash we refer to these as \emph{``big brother boxes"}\textemdash which detect attempts to bypass the intended use of the device. We note that, circumventions in such scenarios are reminiscent of so-called side-channel attacks in quantum key distribution, exploiting weaknesses in the physical implementation of security protocols (see, e.g., Refs.~\cite{LydersenWiechersWittmannElserSkaarMakarov2010,GerhardtLiuLamasLinaresSkaarKurtsieferMakarov2011,GerhardtLiuLamasLinaresSkaarScaraniMakarovKurtsiefer2011}).

\section{The controlled switch}
\begin{figure}[h!]
\includegraphics[width=0.43\textwidth]{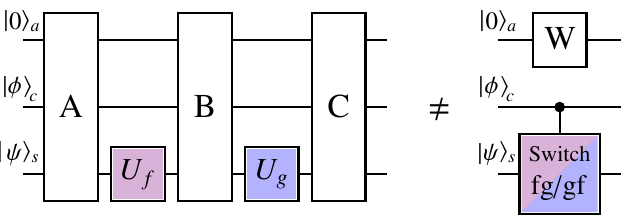}
\caption{\label{fig:controlled order no go}
\textbf{Circuit representation of no-go theorem}. If only single uses of the unknown gates~$U_{f}$ and~$U_{g}$ are permitted, it is not possible to deterministically realize a generic circuit, i.e., with gates $A$, $B$, and $C$ that act on the system and are independent of~$U_{f/g}$, which performs a controlled switch of the operations~$U_{f}$ and~$U_{g}$, that is, for which the order of the operations depends on the control qubit $\ketsub{\phi}{c}$ (see Ref.~\cite{ChiribellaDArianoPerinottiValiron2013}).\vspace*{-3mm}}
\end{figure}

We now return to the second interesting no-go result proven in Ref.~\cite{ChiribellaDArianoPerinottiValiron2013}, the controlled switch of operations (see Fig.~\ref{fig:controlled order no go}). No quantum circuit can be constructed generically to allow deterministic quantum or classical control over the order in which unknown operations are applied, if each operation can be called only once. If the mathematical representations of the operations~$U_{f}$ and~$U_{g}$ may only be inserted once within a sequence of operations that do not depend on~$U_{f}$ and~$U_{g}$, the theorem applies. We can formally write the circuit diagram in Fig.~\ref{fig:controlled order no go} as
\begin{align}
    C_{acs}\,(U_{g})_{s}\,B_{acs}\,(U_{f})_{s}\,A_{acs} &\neq\,W_{a}\bigl(U_{g}U_{f}\oplus U_{f}U_{g}\bigr)_{cs}\,,
\end{align}
with respect to a chosen basis of the control Hilbert space~$\mathcal{H}_{c}$, where we have omitted the tensor product symbols, and~$W$ is an arbitrary operation on the ancilla. This statement is independent of the circuit model of quantum computation. It is a statement that holds\\

\noindent
within quantum mechanics, and its violation would entail drastic consequences, such as the capability ``\emph{to send qubits back in time}"~\cite{ChiribellaDArianoPerinottiValiron2013}.\\
\vspace*{-1mm}

On the other hand, we may interpret the single ``use" of the operations in the more practical sense of being granted limited access to the physical devices implementing the transformations~$U_{f}$ and~$U_{g}$. For instance, an external agent may send two individual laser pulses on an ion. In a photonic setting one may also stay true to the spirit of generic, pre-fabricated setups, by requiring that single copies of the unknown devices are inserted, and that hypothetical photon counters added to these devices would never detect more than one photon.

\section{Implementing $\operatorname{ctrl}$\textendash switch}
Within the mentioned restrictions we now present two practical implementations of the ``movable wires" presented in Ref.~\cite{ChiribellaDArianoPerinottiValiron2013}. In Fig.~\ref{fig:cheating scheme controlled order} we propose a scheme with photons that works on the same premises as the circumvention shown in Fig.~\ref{fig:cheating scheme control U}. A single photon is routed through a setup of beam splitters and half-wave plates where the devices realizing~$U_{f}$ and~$U_{g}$ feature exactly once each. For a similar scheme with ions, on the other hand, all necessary tools are already provided in Fig.~\ref{fig:control with ions}. To enact quantum control over two laser pulses realizing~$U_{f}$ and~$U_{g}$ we follow steps (\ref{step i})\textendash(\ref{step iii}) as before, where the pulse in~(\ref{step iii}) now realizes~$U_{g}$. In addition, we use a mirror to reflect the laser pulses for~$U_{f}$ and~$U_{g}$ back at ion~$2$ at convenient later times. In the next step we apply two $\sigma_{x}$-like pulses, realizing~$S_{g}=\ket{g}\!\bra{g\pr}+\ket{g\pr}\!\bra{g}$ and~$S_{e}=\ket{e}\!\bra{e\pr}+\ket{e\pr}\!\bra{e}$ in Fig.~\ref{fig:control with ions}, tuned to the transitions between~$\ket{g}_{2}$ and~$\ket{g\pr}_{2}$, as well as $\ket{e}_{2}$ and~$\ket{e\pr}_{2}$, respectively, in the subspace of the vibrational ground state~$\ket{0}$. This exchanges the states~$U_{g}\psi$ and~$\psi$. The latter is now stored in the levels~$\ket{g}_{1}\hspace*{-1pt}\ket{0}$ and~$\ket{e}_{1}\hspace*{-1pt}\ket{0}$, and the transition between them is receptive for the second laser pulse~$U_{f}$, followed by the returning pulse~$U_{g}$ (see Fig.~\ref{fig:control with ions}). Consecutively, the $\sigma_{x}$ pulses are applied in time before the reflected pulse~$U_{f}$ acts on the state~$U_{g}\psi$. We finish with steps~(\ref{step iv}) and~(\ref{step v}) from the previous scheme to obtain $\bigl(\alpha\ket{g}_{1}U_{g}U_{f}\ket{\psi}_{2}+\beta\ket{e}_{1}U_{f}U_{g}\ket{\psi}_{2}\bigr)\hspace*{-1pt}\ket{0}$. Summarizing this protocol, the sequence of required steps is: (\ref{step i}), (\ref{step ii}), $U_{g}$, ($S_{g}$,~$S_{e}$), $U_{f}$, $U_{g}$~(returning), ($S_{g}$,~$S_{e}$), $U_{f}$~(returning), (\ref{step iv}), and (\ref{step v}). Note that, although we assume that the pulses~$U_{f}$ and~$U_{g}$ pass through the ion twice, as far as an external agent controlling the pulses is concerned, we have ``used" each operation only once. Only single pulses were sent and each of them has supplied the energy for a single electron transition. The identification of these schemes as realizations of the movable wires depends on the definition of the single use of the operations. For instance, if each interaction of the single laser pulses with the ions is counted as one use, then our scheme may be thought of as realizing the circuit~(18) from Ref.~\cite{ChiribellaDArianoPerinottiValiron2013}, rather than the movable wires.\\

\vspace*{-0.5mm}
\section{Conclusion}
We have supplemented the existing scheme for the control of unknown devices in optical setups~\cite{ZhouRalphKalasuwanZhangPeruzzoLanyonOBrien2011,AraujoFeixCostaBrukner2013} with a feasible ion trap setup based on current technology. We have further presented extensions of both schemes to practically enable the control of the ordering of unknown operations, providing the first specific implementations of\\
\vspace*{-2mm}
\begin{widetext}
\begin{center}
\begin{figure}[hb!]
\includegraphics[width=0.85\textwidth]{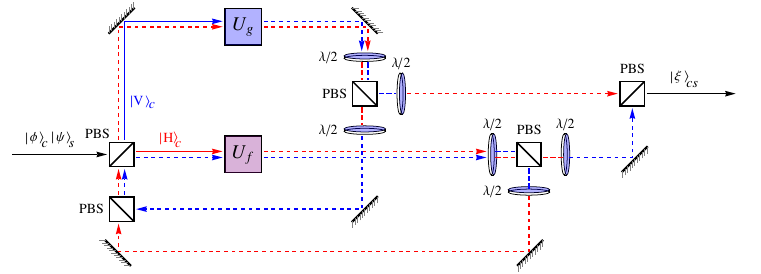}
\caption{\label{fig:cheating scheme controlled order}
\textbf{Quantum control of sequences}. Our scheme enables the quantum control of the ordering in which single devices, labeled by~$U_{f}$ and~$U_{g}$, are applied to the system state~$\ketsub{\psi}{s}$. As before, a single photon in the state $\ketsub{\phi}{c}\ketsub{\psi}{s}$, where the polarization state $\ketsub{\phi}{c}=\alpha\ketsub{\mathrm{H}}{c}+\beta\ketsub{\mathrm{V}}{c}$ serves as the control qubit and $\ketsub{\psi}{s}$ encodes an additional degree of freedom, such as orbital angular momentum, serves as the input. The PBSs are configured to transmit and reflect horizontal and vertical polarization, respectively. The half-wave plates $(\lambda/2)$ are configured to exchange vertical and horizontal polarization, while leaving the system state~$\ketsub{\psi}{s}$ invariant. The unknown devices~$U_{f}$ and~$U_{g}$, on the other hand, are assumed to act on the system only. The beams are recombined at the rightmost PBS, where the final state is $\ketsub{\xi}{cs}=\alpha\ketsub{\mathrm{H}}{c}U_{g}U_{f}\ketsub{\psi}{s}+
\beta\ketsub{\mathrm{V}}{c}U_{f}U_{g}\ketsub{\psi}{s}$.}\vspace*{-10mm}
\end{figure}
\end{center}
\end{widetext}
\noindent
the ``movable wires" conceived in Ref.~\cite{ChiribellaDArianoPerinottiValiron2013} to circumvent the corresponding no-go result. Using these examples as guidance, we have argued that similar features can, in principle, be found in any setup with additionally available dimensions or degrees of freedom, which can be interpreted as generalized interferometric settings. It is interesting to note that the typical method to practically obtain qubits in the first place is to restrict the available state space of the underlying physical system to a two-dimensional subspace.\\

Based on this discussion it appears that the modularity that is desired in certain quantum computational tasks~\cite{ThompsonGuModiVedral2013} can typically be provided in practical situations. Nonetheless, the no-go theorems proven in Refs.~\cite{ChiribellaDArianoPerinottiValiron2013,AraujoFeixCostaBrukner2013,ThompsonGuModiVedral2013}, we believe, are of great significance to adversarial settings, where one party tries to deny flexible use of some device with an unknown function. In this case the provider of the black-box operation aims at enforcing the restrictions of the no-go theorems. For the particular optical circumvention scheme described earlier we have proposed that the provider may add photon number counters to the black-box devices to achieve this. While this prevents the specific schemes shown in Figs.~\ref{fig:cheating scheme control U} and~\ref{fig:cheating scheme controlled order} from functioning as intended, open questions that remain concern the possible strategies for the provider and his or her adversary in general. Moreover, little is known about approximating controlled circuits probabilistically when exact circumventions are denied~\cite{NakayamaSoedaMurao2013}.


\begin{acknowledgements}
\vspace*{-4mm}
We are grateful to Tom\'{a}\v{s} Ryb\'{a}r, as well as Mateus Ara\'{u}jo, \v{C}aslav Brukner, Fabio Costa, Adrien Feix,  Petar Jurcevic, Benjamin~P. Lanyon, and Philip Walther, for valuable discussions and comments. We acknowledge funding by the Austrian Science Fund (FWF) through the SFB FoQuS: F4012.
\end{acknowledgements}


\end{document}